\documentstyle[12pt]{article}

         \def\tr{{\rm tr}}
         \def\la{\lambda}
         \def\be{\begin{equation}}
         \def\bea{\begin{eqnarray}}
         
         \def\o{\over}

         \def\ee{\end{equation}}
         \def\eea{\end{eqnarray}}
         \def\R{\rm {I\kern-.200em R}}
         \def\C{\rm {I\kern-.520em C}}
\begin{document}
\begin{titlepage}
\begin{center}
{ \Large \bf
        Phase structure of the generalized two dimensional}
\vskip 7mm
{ \Large \bf
       Yang-Mills theories on sphere}

\vskip 15 mm
{ \bf
M. Alimohammadi$^a$\footnote {e-mail:alimohmd@netware2.ipm.ac.ir},
A. Tofighi $^{b,c} $ \footnote {e-mail:umcc@dci.iran.com}}
\vskip 10 mm
{\it

 $^a$ Department of Physics, University of Tehran,
             North-Kargar Ave.,\\
             Tehran, Iran \\

 $^b$ Department of Physics, University of Mazandaran,
              Babolsar, Iran \\

 $^c$ Institute for Studies in Theoretical Physics and Mathematics, \\
 P.O.Box 19395-5746, Tehran, Iran \\ }

\end{center}
\vskip 3cm
\begin{abstract}
We find a general expression for the free energy of
$G(\phi )=\phi^{2k}$ generalized 2D Yang-Mills theories in the
strong ($A>A_c$) region at large $N$. We also show that in this region,
the density function of Young tableau of these models
is a three-cut problem. In the specific $\phi^6$ model, we
show that the theory has a third order phase transition, like
$\phi^2$ (YM$_2$) and $\phi^4$ models. We have some comments
for $k \geq 4$ cases. At the end, we study the phase structure of
$\phi^2 + g \phi^4$ model for $ g \leq {A \over 4}$ region.

\end{abstract}

\end{titlepage}
\newpage
\section { Introduction }

The pure 2D Yang-Mills  theory (YM$_2$) is defined by the Lagrangian
$\tr(F^2)$ on a compact
Riemann surface. In an equivalent formulation of this theory,
one can use $i\tr(BF)+\tr(B^2)$ as the Lagrangian, where $B$ is an auxiliary
pseudo-scalar field in the adjoint representation of the gauge group. Path
integration over field $B$ leaves an effective Lagrangian of the form $\tr(F^2)$.

Pure YM$_2$ theory on a compact Riemann surface is characterized by its invariance
under area preserving diffeomorphism and by the fact that there are no propagating
degrees of freedom. Now it is easy to realize that these properties are not unique
to $i\tr(BF)+\tr(B^2)$ theory, but rather are shared by a wide class of theories, called
the generalized Yang-Mills theories (gYM$_2$). These theories are defined by
replacing the $\tr(B^2)$ term by an arbitrary class function $\Lambda (B)$ [10].
Besides the reasons which was discussed in [1] to justify the study of gYM$_2$,
we can add two other reasons. The first one is that the Wilson loop vacuum
expectation value of gYM$_2$ obeys the famous area law behaviour, [11],
and this behaviour is
a signal of confinement which is one of the most important unsolved problem
of QCD. Second, the existance of the third order phase transition in some of the
gYM$_2$ theories (one case was studied in [5] and other examples will be studied
in this paper), is another indication for equivalence of YM$_2$ and gYM$_2$  as
a 2D counterpart of the theory of strong interaction.

In ref.[10], the partition function of gYM$_2$ has been calculated by regarding
the generalized Yang-Mills action as perturbation of topological theory at zero
area. In ref.[1], the partition function has been obtained by generalization
the Migdal's suggestion about the local factor of plaquettes, and has been
shown that this generalizatin satisfies the necessary requirements. And in ref.[11],
this function has been calculated in continuum approach and by standard path
integral method.
The gYM$_2$ theories can be further coupled to fermions, thus obtaining QCD$_2$ and
generalized QCD$_2$ theories [1]. Recently there has been a lot of interest in
these theories. Phase structure, string interpretation and algebraic aspects of
these theories are reviewed in [2].

In this article we explore the issue of phase structure of the gYM$_2$
theories. An early study of phase transition of YM$_2$ in the large-$N$ limit
on the lattice [3] reveals a third order phase transition. The study of
pure continuum YM$_2$ in the large-$N$ limit on the sphere also yields a
similar result [4]. This result is obtained by calculating the free energy
as a function of the area of the sphere ($A$) and distinguishing between the
small and large area behaviour of this function. In [5], the authors considered
gYM$_2$ in the large-$N$ limit on the sphere and they found an exact expression
for an arbitrary gYM$_2$ theory in the weak ($A<A_c$) region, where $A_c$ is
the critical area. They also have found a third order phase transition for the
specific $\phi^4$ model.

In this paper we discuss the issue of phase transition for a wider class of
theories. In section 2 we review the derivation of the free energy and the
density function in the weak ($A<A_c$) region. In section 3 we study the
$\phi^{2k}$ theories. First, we show that the density function of these models
has two maxima in the ($A<A_c$) region, similar to $\phi^4$ model. This, then
enable us to use the method of ref. [5] to obtain a general expression for
the free energy for these theories in the strong ($A>A_c$) region. In section
4 we compute the free energy near the transition point for the specific
$\phi^6$ model and we show this model also has a third order pahse transition.
We also make some remarks for $k \geq 4$ models. And finally in section 5 we
study another class of models, namely the $ \phi^2 + g \phi^4 $ models. If
$g \leq {A\over 4}$ then the density function will have only one maximum at the
origin. We show that these models also undergo a third order phase transition in
this domain.

\section { Large-$N$ behaviour of gYM$_2$ at $A<A_c$ }

The partition function of the gYM$_2$ on a sphere is [5, and references
there-in]
\be Z=\sum_r d_r^2e^{-A\Lambda (r)}, \ee
where $r$'s label the irreducible representations of the gauge group, $d_r$
is the dimension of the $r$-th representation, $A$ is the area of the sphere
and $\Lambda (r) $ is :
\be \Lambda (r) =\sum_{k=1}^p {a_k \over {N^{k-1}}}C_k(r), \ee
in which $C_k$ is the $k$-th Casimir of group, and $a_k$'s are arbitrary
constants. Now consider the gauge group
$U(N)$ and parametrize its representation by $n_1\geq n_2\geq\cdot \cdot
\cdot\geq n_N$, where $n_i$ is the length of the $i$-th row of the Young
tableau. It is found that
$$d_r=\prod_{1\leq i < j \leq N } (1+{{n_i-n_j}\over {j-i}}) $$
\be C_k=\sum_{i=1}^N[(n_i+N-i)^k-(N-i)^k].\ee
To make the partition function (1) convergent, it is necessary that $p$ in
eq.(2) be even and $a_p >0$.

Now, follwing [4], one can write the partition function (1) at large $N$, as
a path integral over continuous parameters. We introduce the continuous
function:
\be \phi (x)=-n(x)-1+x, \ee
where
\be 0 \leq x:=i/N \leq 1 \ \ \ \ {\rm and } \ \ \ \ n(x):=n_i/N. \ee
The partition function (1) then becomes
\be Z=\int \prod_{0\leq x \leq 1} d\phi(x)e^{S[\phi(x)]},\ee
where
\be S(\phi)=N^2\{-A\int^1_0dxG[\phi(x)]+\int_0^1dx\int_0^1dy \ \
{\rm log}|\phi (x)-\phi (y)| \ \ \},\ee
apart from an unimportant constant, and
\be G[\phi ]=\sum_{k=1}^p(-1)^ka_k\phi^k . \ee
Now we introduce the density
\be \rho [\phi (x)]={dx \over d\phi (x)},\ee
where in the cases which $G$ is an even function, the normalization condition for $\rho$ is
\be\int^a_{-a}\rho (\la )d \la=1 .\ee
The function $\rho(z)$ in this case is [5]
\be \rho (z)={\sqrt{a^2-z^2}\over \pi}\sum_{n,q=0}^\infty
{(2n-1)!!\over 2^nn!(2n+q+1)!}a^{2n}z^qg^{(2n+q+1)}(0), \ee
where
\be g(\phi )={A\over 2} G'(\phi ), \ee
and $g^{(n)}$ is the $n$-th derivative of $g$. Similarly one can express the
normalization condition, eq.(10), as
\be \sum_{n=0}^\infty {(2n-1)!!\over 2^nn!(2n-1)!}
    a^{2n}g^{(2n-1)}(0)=1.\ee
Defining the free energy as
\be F:=-{1\over N^2}{\rm ln}Z, \ee
then the derivative of this free energy with respect to the area of the
sphere is
\be F'(A)=\int_0^1dx \ \ G[\phi (x)]=\int_{-a}^ad\la \ \ G(\la)\rho (\la) . \ee
The condition $n_1\geq n_2\geq\cdot \cdot \cdot\geq n_N$ imposes the following
condition on the density $ \rho (\la )$
\be \rho (\la ) \leq1 . \ee
Thus, first we determine the parameter $a$ from eq. (13), then we calculate
$F'(A)$ from eq. (15). Note that the above solution is valid in the weak ($A<A_c$)
region, where $A_c$ is the critical area. If $A>A_c$, then the condition
$\rho \leq 1$ is violated.

\section { The $G(\phi )=\phi^{2k}$ model }

In order to study the behaviour of any model in the strong ($A>A_c$) region, we
need to know the explicit form of the density $\rho$ in the weak ($A<A_c$)
region. From eq. (11) we can obtain $\rho$ for any even function $G(\phi )$.
However, in this section we consider a simple case, namely the
$G(\phi )=\phi^{2k}$ model with arbitrary positive integer $k$ .
The density function $\rho$ in the weak
region is
\be \rho (z)={kA \over \pi}{\sqrt{a^2-z^2}}\sum_{n=0}^{k-1}
{(2n-1)!!\over 2^nn!}a^{2n}z^{2k-2n-2}. \ee
The interesting point is that the above density function has only one minimum
at $z=0$, and two maxima which are symmetric with respect to the origin. To
see this, notice that equating the derivative of $\rho$ equal to zero, will yield $z=0$, and
\be f(y)=-1+\sum_{n=0}^{k-2}{(2n-1)!!\over 2^{n+1}(n+1)!}y^{-(n+1)}=0, \ee
where $y=z^2/a^2$.
Now as all of the coefficients of $y$, in eq. (18), are positive, the function $f(y)$
is a monotonically decreasing function
and has only one root,{\it i.e.} $y_0$. Next, expanding $\rho$ near the
origin, we obtain
\be \rho (z)={kA \over \pi}{(2k-3)!!\over 2^{k-1}(k-1)!}a^{2k}
(1+{2k-1\over 2(2k-3)}z^2a^{-2}+\cdot \cdot \cdot), \ee
thus, $\rho'' (0)>0$. The origin is then a minimum. But, near the points
$z= \pm a$ the curve $\rho (z)$ is concave downward, consequently the points
$z_0^{(\pm)}=\pm a\sqrt{y_0}$ will correspond to two symmetric maxima of the
density.
Therefore in the strong region, all $\phi^{2k}$'s models
are {\it three-cut} problems.
The function $F'(A)$ for $G(\phi )=\phi^{2k}$ in the weak regoin is [5]
\be F_w'(A)={1 \over 2kA}. \ee

Next, we study the strong ($A>A_c$) region. Following [5], we use the following ansatz for
$\rho $
\be
\rho_s(z)=\cases{ 1,&$z\in [-b,-c]\bigcup
[c,b]=:L'$  \cr
\tilde{\rho}_s(z),&$ z\in [-a,-b]\bigcup [-c,c] \bigcup [b,a]=:L.$\cr
}
\ee
Then, if we define the function $H_s(z)$ [6]
\be H_s(z):={\rm P}\int_{-a}^adw \ \ {\rho_s(w)\over z-w}, \ee
where, P indicates the principal value of the integral, it has the follwing
expansion at large z
\be H_s(z)={1\o z}+{1\o z^3} \int_{-a}^a\rho_s(\la)\la^2d\la
+\cdot\cdot\cdot+{1\o z^{2k+1}}F_s'(A)+\cdot\cdot\cdot.  \ee
Hence, one can obtain $F_s'(A)$ via expansion of $H_s(z)$.

To calculate the function $H_s(z)$, we repeat the same steps which was
followed in [5], and using some complex analysis techniques ([7]), which
yield the following result for $\phi^{2k}$ model:
\be H_s(z)=kAz^{2k-1}-kAR(z)\sum_{n,p,q=0}^\infty {\alpha (n,p,q)
z^{2(k-n-p-q-2)}}-2R(z)\int_c^b {\la d\la \o (z^2-\la^2)R(\la)}. \ee
where
\be \alpha (n,p,q)={(2n-1)!!(2p-1)!!(2q-1)!!\over 2^{n+p+q}n!p!q!}
a^{2n}b^{2p}c^{2q},  \ee
and
\be  R(z)=\sqrt{(z^2-a^2)(z^2-b^2)(z^2-c^2)}.  \ee
Now, we expand $H_s(z)/R(z)$ and demand that it behaves like $1/z^4$ at
large $z$. It can be shown that the coefficients of positive powers of $z$
in the above expansion are identically zero. Next, we calculate the
coefficients of $1/z^2$ this gives
\be kA\sum_{n,p,q=0}^\infty {\alpha (n,p,q)} -2\int_c^b {\la d\la \o R(\la)}=0,  \ee
in which $n+p+q=k-1$. By setting the coefficient of
$1/z^4$ to unity we obtain
\be kA\sum_{n,p,q=0}^\infty {\alpha (n,p,q)}-2\int_c^b {\la^3 d\la \o R(\la)}=1,  \ee
where $n+p+q=k$. In $k=2$ case, $\phi^4$ theory, eqs.(27) and (28) reduces to

\be A(a^2+b^2+c^2)=2\int_c^b{\la d\la\o R(\la)},  \ee
and
\be A\left[ \ \ {3\o 4}(a^4+b^4+c^4)+{1\o 2}(a^2b^2+b^2c^2+c^2a^2) \ \ \right]
-2\int_c^b{{\la^3 d\la}\o {R(\la)}}=1,  \ee
which are the same equations that was obtained in [5].

We can express the action in terms of $\rho_s(z)$. If we maximize this
action along with the normalization condition eq.(10), as a constraint, we
obtain another equation. The procedure is fully explained in [5,8]. The
result is
\be kA\sum_{n,p,q=0} \int_c^b \alpha (n,p,q)z^{2(k-n-p-q-2)}R(z)dz
+2\int_c^b dz \ \ {\rm P}\int_c^b \ \
{{R(z)\la d\la} \o{ (z^2-\la^2)R(\la)}} =0. \ee
Note that we can determine three unknown parameters $a$, $b$, and $c$ from
equations (27), (28), and (31).

To compute the function $F_s'(A)$, we start from the function $H_s(z)$
directly. First we expand the $R(z)$
\be R(z)=z^3\sqrt{(1-{a^2 \o z^2})(1-{b^2 \o z^2})(1-{c^2 \o z^2})}=
  -z^3\sum_{n',p',q'=0}^\infty {\beta (n',p',q') z^{-2(n'+p'+q')}},  \ee
where
\be \beta (n,p,q)={(2n-3)!!(2p-3)!!(2q-3)!!\over 2^{n+p+q}n!p!q!}
a^{2n}b^{2p}c^{2q}.  \ee
Furthermore, we define $(-3)!!=-1$. If we substitute the above expansion into
the eq.(24), we obtain
$$ H_s(z)=kAz^{2k-1}+kA\sum_{n,p,q,n',p',q'=0} {\alpha (n,p,q)\beta (n',p',q')
\over z^{2(n+p+q-k)+2(n'+p'+q')+1}} $$
\be +2\sum_{n=0}^\infty
\sum_{n',p',q'=0} {\beta (n',p',q')\over z^{2(n+n'+p'+q')-1}}
\int_c^b{{\la^{2n+1}d\la} \o {R(\la)}}. \ee
From eq.(23) we see that the coefficient of $1/z^{2k+1}$ in the expansion of
$H_s(z)$ is $F_s'(A)$. Therefore from the above equation we get
\be F_s'(A)=kA\sum_{n,p,q,n',p',q'=0} {\alpha (n,p,q) \beta (n',p',q')}
+2\sum_{n,n',p',q'=0} {\beta (n',p',q')}\int_c^b{\la^{2n+1} d\la \o R(\la)}. \ee
In addition, in the first summation of eq. (35) we have the follwing conditions
on the indices:
$$    (n+p+q)+(n'+p'+q')=2k,    \eqno (36-a) $$
and
$$   2k-2n-2p-2q-4 \geq 0.         \eqno (36-b) $$
The condition (36-a) appears due to selection of a specific power of $z$ in the
expansion, and the condition (36-b) is due to complex integration. Furthermore in
the second summation we have the follwing condition on the indices
$$     n+(n'+p'+q')=k+1.         \eqno (36-c)  $$

In this way, we find the explicit relation of the free energy of $\phi^{2k}$'s
models.
For the $k=2$ case, our results
agree with those of ref. [5].

\section { The $G(\phi )=\phi^6$ model }

As an application of the previous results, we will investigate carefully the
phase structure of the $G(\phi )=\phi^6$ model. From eq.(17) we obtain the
density $\rho $ for this model in the weak region, the result is
\setcounter{equation}{36}
\be  \rho (z)={3A\o \pi}({3a^4\over 8}+{a^2z^2\over 2}+z^4)\sqrt{a^2-z^2}. \ee
From the normalization condition ( eq.(13)) we obtain $a=(16/{(15A)})^{1/6}$.
In addition from eq.(20) $F_w'(A)=1/{(6A)}$. This density function has a minimum
at z=0 and two maxima at $z_0^{(\pm)}=\pm(\sqrt{\sqrt{3}+1})a/2$. At
$A=A_c$, the density function (37) is equal to
unity at $z_0^{(\pm)}$. Using this fact, we find the critical area $A_c$
\be   A_c=\pi^6( {3125\o 10368}-{{15625\sqrt{3}}\o {93312}}).  \ee
For $A>A_c$ the eq.(37) is not valid.

Next, we analyse this model in the strong ($A>A_c$) region. The equations
(27), (28), and (31) in this case become
\be 3A\left[ \ \ {3\o 8}(a^4+b^4+c^4)+{1\o 4}(a^2b^2+b^2c^2+c^2a^2) \ \ \right]
-2\int_c^b{\la d\la\o R(\la)}=0,  \ee
$$ 3A \Big[ \ \ {5\o 16}(a^6+b^6+c^6)+{3\o 16}(a^2b^4+a^2c^4+b^2a^4+b^2c^4+
c^2a^4+c^2b^4)+{1\o 8}a^2b^2c^2 \ \ \Big]  $$
\be -2\int_c^b{\la^3 d\la\o R(\la)}=1,  \ee
and
\be 3A\int_c^b(z^2+{{a^2+b^2+c^2}\o 2})R(z)dz +2\int_c^b dz \ \ {\rm P}
 \int_c^b \ \ {R(z)\la d\la \o (z^2-\la^2)R(\la)}=0.  \ee

To study the structure of the phase transition, we must consider the theory
on the sphere with
$A=A_c+\epsilon $ area, where $\epsilon$ is an infinitesimal positive number.
In this region, following [5], we use the follwing change of variables
$$ c=s(1-y), $$
\be b=s(1+y), \ee
$$ a=s \sqrt{2\sqrt{3}-2+e}. $$
Note that these parameters are introduced so that at critical point,
$e$ and $y$ are equal to zero
and $s$ is equal to $z_0^+  $. Now expanding the equations (39), (40) and (41),
we find
$$ As^4(18-6\sqrt{3})-{\pi \o \eta s}+(As^4({9\sqrt{3}\o 2}-3)+{\pi \o \eta s}({1\o 2}
+{\sqrt{3}\o 3}))e  $$
$$ + ({9As^4\o 8}-{\pi \o \eta s}({7\o 8}+{\sqrt{3}\o 2}))e^2+((9+3\sqrt{3})As^4
-{\pi \o \eta s}({9\o 4}+{{4\sqrt{3}}\o {3}}))y^2  $$
\be +({3As^4\o 2}+{\pi \o \eta s}({77\sqrt{3}\o 12}+{89\o 8}))ey^2+(3As^4-{\pi\o \eta s}
({5363 \o 192}+{129\sqrt{3}\o 8}))y^4=0,  \ee
\pagebreak
and
$$ (39\sqrt{3}-57)As^6-1-{\pi s\o \eta }+((42-18\sqrt{3})As^6+{\pi s\o \eta }({1\o 2}+
{{\sqrt{3}}\o 3}))e+(({{45\sqrt{3}}\o{ 8}}-{9\o 2})As^6  $$
$$ -{{\pi s}\o \eta }({7\o 8}+{\sqrt{3}\o 2}))e^2+((33+3\sqrt{3})As^6-{{\pi s}\o \eta }
(2\sqrt{3}+{17\o 4}))y^2+(({3\o 2}+{{9\sqrt{3}}\o 2})As^6  $$
\be +{\pi s\o \eta }({{35\sqrt{3}}\o 4}+{121\o 8}))ey^2+((3\sqrt{3}+24)As^6
-{\pi s\o \eta }({{199\sqrt{3}}\o 8}+{8267\o 192}))y^4=0,  \ee
and
$$ 3(1+\sqrt{3})As^5-{\pi\o \eta }({1\o 2}+{\sqrt{3}\o 3})+(({5\sqrt{3}\o 2}+6)
As^5+{\pi\o \eta }({7\o 6}+{2\sqrt{3}\o 3}))e-(({7\sqrt{3}\o 8}+{13\o 8})As^5  $$
$$ +{\pi\o \eta }({5\o 2}+{13\sqrt{3}\o 9}))e^2-((2+4\sqrt{3})As^5+{\pi\o \eta }
({91\sqrt{3}\o 36}+{211\o 48}))y^2+(({25\o 2}+{15\sqrt{3}\o 2})As^5  $$
\be +{\pi\o \eta }({635\o 24}+{275\sqrt{3}\o 18}))ey^2-(({56\o 3}+{32\sqrt{3}\o 3})
+{\pi\o \eta }({60673\sqrt{3}\o 1728}+{23353\o 384}))y^4=0.  \ee
The parameter $\eta =\sqrt{{2\sqrt{3}-3}}$ is used for the sake of brevity in the
above formulas. We also have kept terms up to order $y^4$ or $e^2$ (we will
show that $e$ is of order $y^2$). Next we obtain $s$ from eq.(43), the result
is
$$ s=({\pi\o A})^{1\o 5}({12+7\sqrt{3}\o 648})^{1\o 10}
\left[ 1-({1+\sqrt{3}\o 8})e
+({5\o 32}+{\sqrt{3}\o 12})e^2 \right. $$
\be \left. +({1\o 4}+{\sqrt{3}\o 6})y^2-({167\o 96}+{95\sqrt{3}\o 96})ey^2
+({911\o 192}+{11\sqrt{3}\o 4})y^4 \right] .  \ee
Substituting this $s$ in eq.(45) will give
\be  e=({{5\sqrt{3}}\o 2}-{1\o 2})y^2-({15\o 16}+{{37\sqrt{3}}\o 48})y^4.  \ee
So $e$ is of order $y^2$. Using eq.(44), we obtain
\be    y^2=({4\sqrt{3}\o 15}-{2\o 5})\delta+({317\o 360}-{77\sqrt{3}\o 150})
\delta^2,   \ee
and
\be  e=({11\o 5}-{17\sqrt{3}\o 15})\delta+({8533\sqrt{3}\o 3600}
-{14929\o 3600})\delta^2.  \ee
The parameter $\delta $ is the reduced area, {\it i.e. } $\delta={{A-A_c}\o A_c}$.\\
From eq. (35), we find $F_s'(A)$, the result is
\pagebreak
$$ F_s'(A)={{3A}\o 1024}\Big[ \ \ 35(a^{12}+b^{12}+c^{12})-12(a^6b^6+a^6c^6+b^6c^6) $$
$$ +12a^2b^2c^2(a^4b^2+a^4c^2+b^4a^2+b^4c^2+c^4a^2+c^4b^2)+14a^4b^4c^4 $$
$$ -2a^2b^2c^2(a^6+b^6+c^6)-19(a^8b^4+a^8c^4+b^8a^4+b^8c^4+c^8a^4+c^8b^4) $$
$$ -10(a^{10}b^2+a^{10}c^2+b^{10}a^2+b^{10}c^2+c^{10}a^2+c^{10}b^2) \ \ \Big] $$
$$ +\Big[ \ \ {5\o 64}(a^8+b^8+c^8)-{1\o 16}(a^6b^2+a^6c^2+b^6a^2+b^6c^2
+c^6a^2+c^6b^2)  $$
$$ +{1\o 16}a^2b^2c^2(a^2+b^2+c^2)-{1\o 32}(a^4b^4+a^4c^4+b^4c^4) \ \ \Big]
\int_c^b{\la d\la\o R(\la)}  $$
$$ +{1\o 8}\left[ \ \ a^6+b^6+c^6-(a^2b^4+a^2c^4+b^2a^4+b^2c^4+c^2a^4+c^2b^4)
 +2a^2b^2c^2 \ \ \right]\int_c^b{{\la^3 d\la}\o R(\la)} $$
\be +{1\o 4}\left[ \ \ a^4+b^4+c^4-2(a^2b^2+a^2c^2+b^2c^2)
  \ \ \right]\int_c^b{{\la^5 d\la}\o R(\la)}+(a^2+b^2+c^2)
\int_c^b{{\la^7 d\la}\o R(\la)}
-2\int_c^b{{\la^9 d\la}\o R(\la)}.   \ee
To compute $F_s'(A)$, we express the parameters $a$, $b$ and $c$
in terms of $\delta$, and after a lengthy
calculations eq.(50) reduces to
\be  F_s'(A)={1\o 6A}\left[ \ \ 1+({271\o 10800}+{1289\sqrt{3}\o 504000})
 \delta^2+\cdot\cdot\cdot \ \ \right].  \ee
If we compare this with $F_w'(A)$ given in the begining of this section we find
\be F_s'(A)-F_w'(A)=({271\o 64800}+{1289\sqrt{3}\o 3024000}){1\o A_c}
  ({A-A_c\o A_c})^2+ \cdots.  \ee
Therefore, we have a {\it third order} phase transition, which is the same as the
ordinary $YM_2$ [4] and $G (\phi)=\phi^4$ model [5].
In principle one could study other $\phi^{2k}$ models in the same manner. The
$k=4$ case is solvable analytically, but the expressions become too
complicated. For $k \geq 5 $ one can not solve the equations analytically.

\section{\bf The $G(\phi )=\phi^2+g\phi^4 $ model}

So far in our study of the phase transition for $gYM_2$ theories we
considered only those $G(\phi)$ which contained a single term. In ref.[9]
the authors study the phase transition of $gYM_2$ on a closed surface of arbitrary genus
and area $A$. In particular they investigate the $G(\phi)=\phi^2+g\phi^3$
model. However, their treatment is mostly qualitative. In this section we
consider a simple combination of $\phi^2$ and $\phi^4$, namely we study the
$G(\phi )=\phi^2+g\phi^4$ model.

In the weak region we can obtain the density $\rho$ from eq.(10), the
result is
\be  \rho (z)={A\o \pi}{\sqrt{a^2-z^2}}(1+ga^2+2gz^2).  \ee
The above density will have only one maxima at $z=0$, when
\be 3ga^2\le 1. \ee
The normalization condition (13) yields
\be {1\o 2}Aa^2+{3\o 4}gAa^4=1.  \ee
Using (55) the condition (54) reduces to
\be g\le A/4 . \ee
Therefore, if
this condition is satisfied, we will have a two-cut problem in $A>A_c$ areas and
we limit ourselves to this region ( condition (56)).\\
The free energy of this model is ( using (15))
\be  F_w'(A)={1\o 8}a^4A+{5\o 16}ga^6A+{9\o 64}g^2a^8A.  \ee

To study this model in the strong ($A>A_c$) region we use the following ansatz
for $\rho$ [4].
\be
\rho_s(z)=\cases{ 1,&$z\in [-b,b] $  \cr
\tilde{\rho}_s(z),&$ z\in [-a,-b]\bigcup [b,a] $\cr
}
\ee
Using complex analysis [4,5,6] we obtain the function $H_s(z)$ defined through
eq.(22), the result is
\be  H_s(z)=Az+2gAz^3-\sqrt{(z^2-a^2)(z^2-b^2)}\left[ 2gAz+\int_{-b}^b{d\la
\o (z-\la)U(\la)}\right] ,   \ee
where
\be    U(\la)=\sqrt{(a^2-\la^2)(b^2-\la^2)}. \ee
Using eq.(22) it is seen that $H_s(z)$ should behave as ${1\o z}$ for
large z. Therefore the coefficient of $z$ of eq.(59) must be equal to zero
\be    A+gAM-\int_{-b}^b{d\la\o U(\la)}=0, \ee
and the coefficient of $1/z$ must be one
\be {1\o 2}MA+gA({3\o 4}M^2-N)-\int_{-b}^b{\la^2 d\la\o U(\la)}=1.  \ee
In these relations $M=a^2+b^2$ and $N=a^2b^2$. The two unknown parameters
$a$ and $b$ can be determined from these two equations.

Using equations (15), (22) and (59) we obtain the following expression for
the free energy:
$$ F_s'(A)=({1\o 8}M^2-{1\o 2}N)\int_{-b}^b{d\la\o U(\la)}+{1\o 2}M\int_{-b}
^b{\la^2 d\la\o U(\la)}-\int_{-b}^b{\la^4 d\la\o U(\la)}  $$
$$ +g \left[ \ \ (-{1\o 4}MN+{1\o 16}M^3)\int_{-b}^b{d\la\o U(\la)}+({1\o 8}M^2
-{1\o 2}N)\int_{-b}^b{\la^2 d\la\o U(\la)}+{1\o 2}M\int_{-b}^b{\la^4 d\la\o
U(\la)}-\int_{-b}^b{\la^6 d\la\o U(\la)} \right] $$
\be -2gA({1\o 4}MN-{1\o 16}M^3)-2g^2A(-{1\o 8}N^2+{3\o 16}M^2N-{5\o 128}M^4)
 .\ee
In order to investigate the phase transition, we use the following change of
variables
\be    a=a_c(1-h),  \ee
\be  ({b \o a})^2=k.  \ee
The parameter $a_c$ is the value of $a$ at the critical point, and $h$ and $k$
are equal to zero at this point. At the critical point,
the normalization condition (55) becomes
\be {1\o 2}{A_c}{a_c}^2+{3\o 4}g{A_c}{a_c}^4=1. \ee
In addition, as at $A=A_c$, $\rho (0)$ is equal to one, we obtain
\be {a_c}{A_c}+g{A_c}{a_c}^3=\pi.  \ee
In order to study the phase transition of this model, we limit ourselves to the
small values of $g$ ($g<<1$).
In this limit $A_c$ and $a_c$ become (by eqs.(66) and (67))
\be   A_c={\pi^2\o 2}-g+{4g^2\o \pi^2}+\cdots ,  \ee
and
\be   a_c={2\o \pi}-{4g\o \pi^3}+{40g^2\o \pi^5}+\cdots . \ee
Now, we follow the same steps of previous section. First, we expand the
equations (61) and (62) in terms of $h$ and $k$ and then solve these equations
for $h$ and $k$. The results, up to order $\delta^2$, are
\be h=({1 \o 2} -{3 \o 2}{g \o A_c}+{27 \o 2}{g^2 \o A_c^2})\delta
  +(-{5 \o 8} +{7 \o 4}{g \o A_c}-{287 \o 8}{g^2 \o A_c^2})\delta^2+\cdots , \ee
\be k=(2+14{g \o A_c}+42{g^2 \o A_c^2})\delta
  +(-{7 \o 4} -{93 \o 2}{g \o A_c}+{283 \o 4}{g^2 \o A_c^2})\delta^2+\cdots . \ee
Second, we find $F_s'(A)$ from eq.(63). The result is
\be F_s'(A)={1 \o A}\{ {1\o 2}-{1\o 2}{g\o A}+{9\o 4}{g^2\o A^2}+(1+3{g\o A}
+85{g^2\o A^2})\delta^2+\cdots \}. \ee
It is understood that this relation is true only up to order $g^2$ and $\delta^2$.
To compare the equation (72) with $F'(A)$ in the weak region, we first note that
$a^2$, up to order $g^2$, is (from (55))
\be a^2={2\o A}(1-3{g\o A}+18{g^2\o A^2}), \ee
and from this, $F_w'(A)$ becomes (from (57))
\be F_w'(A)={1\o A}({1\o 2}-{1\o 2}{g\o A}+{9\o 4}{g^2\o A^2}). \ee
Therefore
\be  F_s'(A)-F_w'(A)={1\o A_c}(1+3{g\o A_c}+85{g^2\o A_c^2})({{A-A_c}\o A_c})^2+\cdots .  \ee
This shows that the $G(\phi )=\phi^2+g\phi^4$ model in $g<A/4$ region has also a
{\it third order} phase transition. Also note that at $g=0$, the eq.(75) reduces
to
\be  F_s'(A)-F_w'(A)={2\o \pi^2}({{A-A_c}\o A_c})^2+\cdots ,  \ee
which is the same relation that was obtained in ref.[4] for ordinary 2-dimensional
Yang-Mills theory.

At the end we should mention that for $g>A/4$, the density function $\rho$
(eq.(53)) has two symmetric maxima in the weak ($A<A_c$) region. So we have
a three-cut problem and the above perturbative calculation is not possible.
\vskip 1cm
\noindent{\bf Acknowledgement}

We would like to thank M. Khorrami for useful discussion. M. A. would also
like to thank the Institute for Studies in Theoretical Physics and Mathematics,
and research council of University of Tehran for partial financial support.

\end{document}